\documentclass[sigconf]{acmart}

\copyrightyear{2025}
\acmYear{2025}

\usepackage{multirow}
\usepackage{subfigure}
\usepackage{tabularx}   
\usepackage{xcolor}
\usepackage{colortbl}   
\usepackage{soul}

\usepackage{balance}

\definecolor{lightgreen}{RGB}{173, 216, 230}
\definecolor{lightred}{RGB}{255, 218, 185}
\definecolor{MyLightRed}{RGB}{255,150,150} 
\definecolor{MyDarkRed}{RGB}{139,0,0}     
\begin{document}

\def\method{\textsc{Debater}}



\title{Learning Refined Document Representations for Dense Retrieval via Deliberate Thinking}

\author{Yifan Ji}
\authornote{ indicates equal contribution.}
\affiliation{%
  \institution{Northeastern University}
  \city{Shenyang}
  \country{China}
}
\email{bigtailwolf001@gmail.com}

\author{Zhipeng Xu}
\authornotemark[1]
\affiliation{%
  \institution{Northeastern University}
  \city{Shenyang}
  \country{China}
}
\email{xuzhipeng@stumail.neu.edu.cn}

\author{Zhenghao Liu}
\authornote{ indicates corresponding author.}
\affiliation{%
  \institution{Northeastern University}
  \city{Shenyang}
  \country{China}
}
\email{liuzhenghao@mail.neu.edu.cn}

\author{Yukun Yan}
\affiliation{%
  \institution{Tsinghua University}
  \city{Beijing}
  \country{China}
}
\email{yanyk.thu@gmail.com}

\author{Shi Yu}
\affiliation{%
  \institution{Tsinghua University}
  \city{Beijing}
  \country{China}
}
\email{yushi17@foxmail.com}

\author{Yishan Li}
\affiliation{%
  \institution{ModelBest Inc.}
  \city{Beijing}
  \country{China}
}
\email{liyishanthu@gmail.com}

\author{Zhiyuan Liu}
\affiliation{%
  \institution{Tsinghua University}
  \city{Beijing}
  \country{China}
}
\email{liuzy@tsinghua.edu.cn}

\author{Yu Gu}
\affiliation{%
  \institution{Northeastern University}
  \city{Shenyang}
  \country{China}
}
\email{guyu@mail.neu.edu.cn}

\author{Ge Yu}
\affiliation{%
  \institution{Northeastern University}
  \city{Shenyang}
  \country{China}
}
\email{yuge@mail.neu.edu.cn}

\author{Maosong Sun}
\affiliation{%
  \institution{Tsinghua University}
  \city{Beijing}
  \country{China}
}
\email{sms@tsinghua.edu.cn}

\renewcommand{\shortauthors}{Yifan Ji et al.}

\begin{abstract}
Recent dense retrievers increasingly leverage the robust text understanding capabilities of Large Language Models (LLMs), encoding queries and documents into a shared embedding space for effective retrieval.
However, most existing methods represent each document with a single embedding, which is less effective at capturing its multifaceted semantics and thereby limits matching accuracy. In this paper, we propose \textbf{D}\textbf{e}li\textbf{b}er\textbf{a}te \textbf{T}hinking based Dens\textbf{e} \textbf{R}etriever (\method{}), a novel approach that enhances document representations by incorporating a step-by-step thinking process. \method{} introduces a Chain-of-Deliberation mechanism, which iteratively refines document embeddings through a continuous chain-of-thought. To integrate information from various thinking steps, \method{} further employs a Self Distillation mechanism that identifies and fuses the most informative steps into a unified embedding. Experimental results show that \method{} significantly outperforms existing methods across several retrieval benchmarks, demonstrating superior accuracy and robustness. All codes and datasets are available at https://github.com/OpenBMB/DEBATER.
\end{abstract}




\maketitle

\section{Introduction}
Dense retrieval models encode both queries and documents into a dense embedding space and measure their similarity to retrieve relevant documents~\cite{karpukhin2020dense, zhao2024dense, xiong2020approximate}, demonstrating strong effectiveness in various downstream NLP tasks, such as open-domain question answering~\cite{chen2020open}, fact verification~\cite{liu2019fine}, and web search~\cite{chen2024ms}. However, recent findings have shown that dense retrievers suffer from significant performance degradation when applied to new tasks or domains~\cite{su2022one}, raising concerns about their versatility~\cite{luo2024large, khramtsova2024leveraging}.

\begin{figure}[t]

\begin{center}

\includegraphics[width=0.95\linewidth]{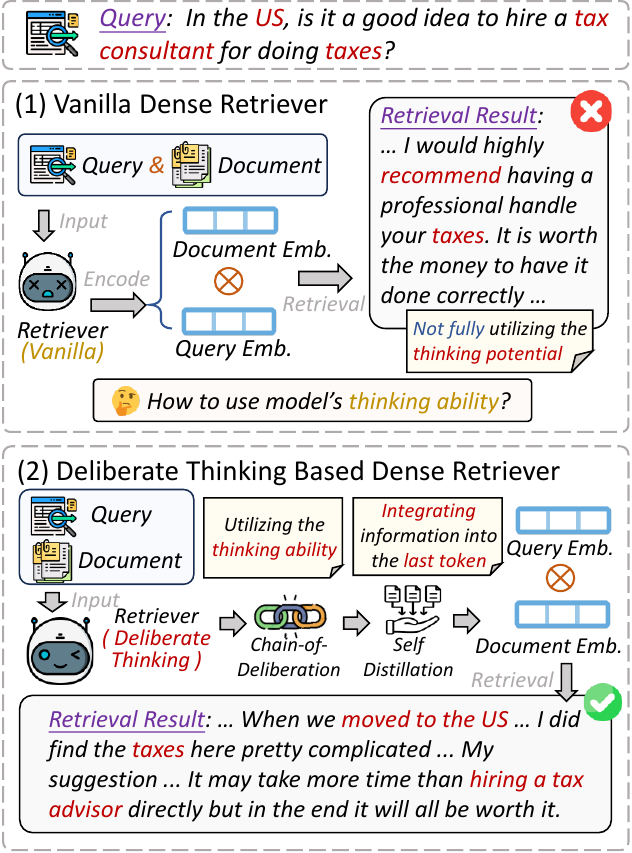}
\end{center}

\caption{The Illustration of Our \textbf{D}\textbf{e}li\textbf{b}er\textbf{a}te \textbf{T}hinking based Dens\textbf{e} \textbf{R}etriever (\method{}). \method{} leverages the reasoning capability of LLM to conduct fine-grained document representations for retrieval.}

\label{fig:intro}

\end{figure}
Large Language Models, such as ChatGPT~\cite{achiam2023gpt} and LLaMA~\cite{touvron2023llama}, have demonstrated extraordinary emergent capabilities~\cite{wei2022emergent, zhao2023survey}, inspiring researchers to leverage them to enhance the task and domain generalization of dense retrievers~\cite{zhu2023large, khramtsova2024leveraging}. In particular, existing work has focused on prompting LLMs to generate dense representations for retrieval~\cite{zhuang2024promptreps}. These methods typically use task-specific instructions or in-context demonstrations to guide LLMs in generating task- and domain-aware embeddings. To learn more tailored representations for dense retrieval, researchers further focus on optimizing LLM-based retrievers using relevance labels~\cite{ma2024fine, neelakantan2022text, li2025making}. These methods exploit the superior reasoning abilities of LLMs, achieving impressive performance across various retrieval tasks~\cite{wang2023improving, zhu2023large, luo2024large}. Recent studies suggest that LLMs pose strong reasoning capability, particularly implemented by their step-by-step thinking~\cite{kudo2024think,wei2022chain}. LLM-based retrievers typically rely on the hidden state of the end-of-sequence token as both query and document representations. Nevertheless, only relying on one embedding usually shows less effectiveness in representing documents from different views that can match queries~\cite{zhang2022multi,khattab2020colbert}. 

In this paper, we propose a \textbf{D}\textbf{e}li\textbf{b}er\textbf{a}te \textbf{T}hinking based Dens\textbf{e} \textbf{R}etriever (\method{}) model to learn more effective document representations through deliberately thinking step-by-step before retrieval. As shown in Figure~\ref{fig:intro}, our method stimulates LLMs to conduct the reasoning process, enabling them to generate more fine-grained document representations for retrieval. Specifically, \method{} introduces the Chain-of-Deliberation mechanism to encourage LLMs to conduct deliberate thinking by autoregressively decoding the document representations. Then \method{} utilizes the Self Distillation mechanisms to gather all information from previous steps and compress them into the document embedding at the last step.

Our experiments demonstrate that \method{} achieves retrieval performance comparable to, or even surpassing, that of baseline methods implemented by larger-scale LLMs, demonstrating its effectiveness. Further analysis reveals that Chain-of-Deliberation and Self-Distillation play complementary roles in \method{}, and that increasing the number of reasoning steps appropriately can benefit LLM-based dense retrieval models. The document representations produced by \method{} are progressively refined through iterative thinking, where each step autoregressively generates intermediate representations. By incorporating Self-Distillation, the model is able to extract different salient information at different reasoning steps and integrate them into a more comprehensive and semantically rich final representation.

\section{Related Work}
Dense retrieval~\cite{karpukhin2020dense, xiong2020approximate, su2022one} has proven effective in various NLP downstream tasks~\cite{liu2019fine, chen2024ms, guu2020retrieval}. However, the versatility of dense retrievers remains a challenge that hinders their progress~\cite{luo2024large, lee2024gecko}, particularly their inability to generate task- and domain-specific embeddings and return suitable results~\cite{su2022one, luo2024large, tao2024llms}. To address this limitation, prior work has focused on conducting fine-grained data curation to fine-tune dense retrievers with multi-task instructions~\cite{su2022one, asai2022task}. However, obtaining high-quality relevance labels can be difficult for training dense retrievers~\cite{yu2022coco, gao2022precise, wang2023improving}.

Recent research has shifted towards using LLMs as the backbone for dense retrievers~\cite{tao2024llms}, thriving on their strong emergence capabilities. Some studies attempt to directly prompt LLMs to generate embeddings for retrieval~\cite{zhuang2024promptreps}. However, prompt-based approaches cannot leverage pre-existing retrieval signals, limiting their effectiveness~\cite{zhu2023large}. In contrast, recent efforts have focused on fine-tuning LLMs for dense retrieval tasks~\cite{wang2023improving, ma2024fine, li2024llama2vec}, or designing additional pretraining tasks to transform LLMs into dense retrievers~\cite{behnamghader2024llm2vec}, achieving strong retrieval performance and generalization capabilities. However, existing methods typically extract the last hidden state of the end-of-sequence token as the dense representation~\cite{ma2024fine, luo2024large}, which is not always effective for fully representing documents from different perspectives to match queries~\cite{zhang2022multi, khattab2020colbert}. The exploration of different document representations, such as leveraging the reasoning ability of LLMs, remains an underexplored area.

\begin{figure*}[t]
    \centering
    \includegraphics[width=0.95\linewidth]{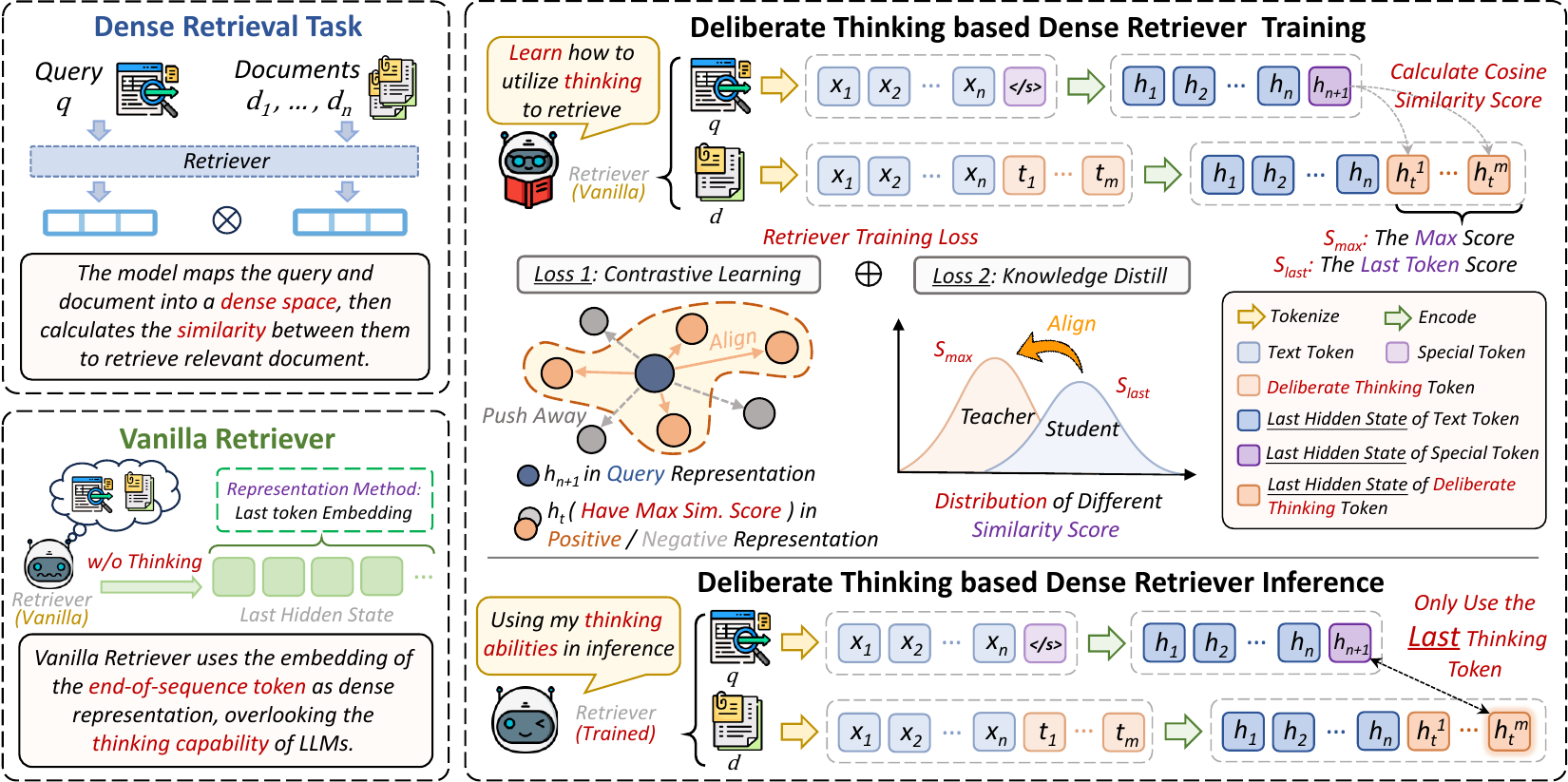}
    \caption{The Architecture of \textbf{D}\textbf{e}li\textbf{b}er\textbf{a}te \textbf{T}hinking based Dens\textbf{e} \textbf{R}etriever (\method{}). 
    }
    \label{fig:main-architecture}
\end{figure*}

To enhance the reasoning capability of LLMs, one approach is to generate intermediate reasoning steps using Chain-of-Thought (CoT)~\cite{wei2022chain} or its variants~\cite{chen2022program, zhang2024wrong}. CoT allows LLMs to delay final answers by engaging in reasoning~\cite{kudo2024think}, improving response accuracy~\cite{wei2022chain, chu2023survey}. However, these approaches operate within the language space and often require generating tens or even hundreds of additional tokens, which can hinder their ability to meet the latency requirements of dense retrievers. Current research is exploring the integration of CoT reasoning into a continuous latent space~\cite{hao2024training, xie2024self} to enhance computational efficiency. Building on these advancements, our \method{} focuses on latent reasoning chains, encouraging LLM-based retrievers to think step-by-step to enhance the dense representations of documents.
\section{Methodology}
In this section, we introduce our Deliberate Thinking based Dense Retriever (\method{}).
We first introduce the preliminary of LLM-based dense retrieval~(Sec.~\ref{method:preliminary}). Then we describe our deliberation thinking based embedding learning method used by \method{}~(Sec.~\ref{method:method}).

\subsection{Preliminary of Dense Retrieval with Large Language Models as Foundations}
\label{method:preliminary}
Given a query $q$ and a document collection $\mathcal{D}$, the goal of the retrieval task is to identify a subset of documents most relevant to the query.

LLM-based dense retrievers typically map both the query $q$ and document $d$ into a shared latent space for retrieval, where the query embedding $h^q$ and document embedding $h^d$ are defined as:
\begin{equation}
\begin{aligned} 
\label{eq:preliminary}
h^q &= \text{LLM}( q, \text{\texttt{</s>}})[-1],\\
h^d &= \text{LLM}( d, \text{\texttt{</s>}})[-1].
\end{aligned} 
\end{equation}
The ranking score $f(q, d)$ between the query embedding $h^q$ and the document embedding $h^d$ is calculated as:
\begin{equation}
f(q, d) = sim(h^q, h^d),
\end{equation}
where \textit{sim} denotes the similarity function. In \method{}, we use cosine similarity to measure the similarity between queries and documents, which is also employed in previous works~\cite{wang2023improving, behnamghader2024llm2vec}. Subsequently, we contrastively train the LLM to maximize the probability of retrieving the positive document $d^+$ over the negative document $d^-$:
\begin{equation}
\begin{aligned} 
\label{eq:prob} 
p(d^{+}|q, {d^{+}}\cup\mathcal{D}^{-}) = \frac{e^{f(q, d^+)}}{e^{f(q, d^+)} + \sum_{d^-\in \mathcal{D}^-} e^{f(q, d^-)}}, \end{aligned} 
\end{equation}
where $\mathcal{D}^{-}$ denotes the set of negative documents, typically obtained via in-batch negative sampling~\cite{karpukhin2020dense}. 

Current LLM-based dense retrievers typically use the last hidden state corresponding to the end-of-sequence token (\texttt{</s>}) as the dense representation. However, they do not fully exploit the reasoning capabilities of LLMs, which helps conduct more effective representations by learning information from diverse views of documents.

\subsection{Enhancing Dense Retriever through Deliberate Thinking}
\label{method:method} 
In this subsection, we introduce the Deliberate Thinking based Dense Retriever (\method{}), which aims to unleash the reasoning ability of LLMs and generate more fine-grained document representations. As shown in Figure~\ref{fig:main-architecture}, \method{} consists of two modules to enhance the LLM-based dense retriever: Chain-of-Deliberation (CoD) and Self Distillation (SD).

\textbf{Chain-of-Deliberation.}
To enhance these LLM-based dense retrievers, \method{} introduces the Chain-of-Deliberation (CoD) approach, which delays the computation of document embeddings by performing several steps of reasoning.

Specifically, CoD incorporates a sequence of prompt tokens $\{t_1, t_2, \dots, t_m \}$ to stimulate the reasoning capability of LLMs when representing the document $d$. These tokens $\{t_1, t_2, \dots, t_{m-1} \}$ serve as intermediate thinking steps, encouraging the model to think step-by-step before producing the final document embedding at the $m$-th step:
\begin{equation}
\label{eq:cod}
h_{m}^d = \text{LLM}( X,t_1,t_2, \dots, t_{m-1}, t_m),
\end{equation}
where $m$ is a hyperparameter to control the thinking depth. An appropriate choice of $m$ is crucial to avoid overthinking or under-optimization.

During training, we first calculate the similarity score between query representation $h^q$ and the document representation $h^d_i$ at the $i$-th thinking step:
\begin{equation}
f(q, d(t_i)) = sim (h^q, h^d_i).
\end{equation}
Next, we gather all similarity scores using the decoded document representations $\{h^d_1, ..., h^d_m\}$. We then select the most useful thinking step from CoD and use the corresponding embedding as the document representation to compute the training loss. The relevance scores $f_\text{max}(q, d)$ between the query and the document are computed as:
\begin{equation} 
f_\text{max}(q, d) = \max_{1 \leq i \leq m} sim(h^q, h^d_i),
\label{eq:relevance}
\end{equation} 
The LLM is optimized by minimizing the contrastive training loss:
\begin{equation}
\mathcal{L}_c=-\text{log}\frac{e^{f_\text{max}(q, {d^+})}}{e^{f_\text{max}(q, {d^+})}+ \sum_{d^-\in \mathcal{D}^-}e^{f_\text{max}(q, {d^-})}}.
\end{equation}

\textbf{Self Distillation.}
Although the final token of the Chain-of-Deliberation aggregates information from all thinking steps through autoregressive decoding, it may overlook crucial reasoning cues presented in embeddings decoded at earlier steps.

To address this, we introduce Self Distillation (SD), a strategy for distilling knowledge from different thinking steps into the final document representation $h^d_m$. Specifically, we use the most informative thinking step as the teacher to guide the representation learning of the final token in CoD, thereby enhancing the document representation.

For the query $q$, we compute the ranking probability of the $i$-th document $d_i$ in the document collection $\Tilde{\mathcal{D}} = \{d^+\} \cup\mathcal{D}^{-}$ as: 
\begin{equation}
P(d_i | q) = \frac{e^{f_\text{max}(q, d_i)}}{\sum_{d_j \in \Tilde{\mathcal{D}}} e^{f_\text{max}(q, d_j)}},
\end{equation}
where $|\Tilde{\mathcal{D}}| = k$. This yields a probability distribution $P(\Tilde{\mathcal{D}}|q)$ over the $k$ documents: 
\begin{equation}
P(\Tilde{\mathcal{D}}|q) = \left[ P(d_1 | q), P(d_2 | q), \dots, P(d_k | q) \right].
\end{equation}
Each value $P(d_i | q)$ represents the ranking probability of the $i$-th document $d_i$ using the document representations from all thinking steps $\{h^d_1, ..., h^d_m\}$ of CoD that yield the highest similarity with the query. Concurrently, we compute the rank probability of $d_i$ using the last-token embedding $h^d_m$ from CoD:
\begin{equation}
Q(d_i(t_m) | q) = \frac{e^{f(q, d_i(t_m))}}{\sum_{d_j \in \{d^+\} \cup\mathcal{D}^{-}} e^{f(q, d_j(t_m))}}.
\end{equation}
Then we can obtain the ranking probability distribution $Q(\Tilde{\mathcal{D}}|q)$ as well:
\begin{equation}
Q(\Tilde{\mathcal{D}}|q) = \left[ Q(d_1 | q), Q(d_2 | q), \dots, Q(d_k | q) \right].
\end{equation}
We then minimize the Kullback-Leibler (KL) divergence between two probability distributions $P(\Tilde{\mathcal{D}}|q)$ and $Q(\Tilde{\mathcal{D}}|q)$:
\begin{equation}
\mathcal{L}_{t}={P(\Tilde{\mathcal{D}}|q) \cdot \log  \frac{P(\Tilde{\mathcal{D}}|q)}{Q(\Tilde{\mathcal{D}}|q)}},
\end{equation}
where the Self Distillation loss $\mathcal{L}_{t}$ optimizes the document representation $h_m^d$ by capturing more crucial matching signals from all thinking steps.

\textbf{Training.}
Finally, we train our \method{} models by minimizing the following loss $\mathcal{L}$:
\begin{equation}
\mathcal{L} = \mathcal{L}_c + \mathcal{L}_t,
\end{equation}
where $\mathcal{L}_c$ optimizes the CoD, and $\mathcal{L}_t$ is used to distill crucial information from the thinking steps into the final dense representation of the document. This combined loss allows \method{} to leverage both thinking depth and self-knowledge distillation to improve retrieval performance.

\section{Experimental Methodology}
In this section, we describe the datasets, evaluation metrics, baselines, and implementation details for our experiments.

\textbf{Datasets.} 
We train all \method{} models using the public portion of the E5 dataset~\cite{wang2023improving,springer2024repetition}, a carefully curated collection of approximately 1.5M publicly available samples. 
Table~\ref{tab:train_dataset_details} presents its statistics, and the full dataset is accessible on their website\footnote{\url{https://github.com/jakespringer/echo-embeddings}}.
\begin{table}[t]
\centering
\caption{Data Statistics of E5 Dataset. We show the composition and distribution of E5 Dataset.}
\begin{tabular}{l|r|r} 
\toprule
\textbf{Dataset} & \textbf{\#Samples} & \textbf{Proportion} \\ \midrule
ELI5 \cite{fan2019eli5longformquestion} & 32,547&2.16\% \\
HotpotQA \cite{yang2018hotpotqadatasetdiverseexplainable} &90,447&5.99\% \\
FEVER \cite{thorne-etal-2018-fever} & 101,578&6.73\% \\
MIRACL \cite{zhang-etal-2023-miracl} & 32,561& 2.16\%\\
MSMARCO Passage Ranking \cite{bajaj2018msmarcohumangenerated} & 249,592&16.53\% \\
MSMARCO Document Ranking \cite{bajaj2018msmarcohumangenerated} & 73,400&4.86\% \\
NQ \cite{kwiatkowski-etal-2019-natural} & 100,231&6.64\% \\
NLI \cite{gao-etal-2021-simcse} & 277,230&18.36\% \\
SQuAD \cite{rajpurkar-etal-2016-squad} & 87,599&5.80\% \\
TriviaQA \cite{joshi2017triviaqa} & 73,346&4.86\% \\
Quora Duplicate Questions \cite{quora-question-pairs} & 101,762&6.74\% \\
Mr-TyDi \cite{zhang-etal-2021-mr} & 48,715&3.23\% \\
DuReader \cite{he-etal-2018-dureader} & 86,395 &5.72\% \\
T2Ranking \cite{10.1145/3539618.3591874} & 154,294&10.22\% \\ \bottomrule
\end{tabular}

\label{tab:train_dataset_details}
\end{table}
\begin{table}[t!]
    \centering
    \caption{Data Statistics of the BEIR Benchmark. We show  the task type, along with the number of queries and passages for each dataset.}
    \begin{tabular}{ l | l | l | r }
       \toprule
           \textbf{Dataset } &\textbf{Task }  & \textbf{\#Query} & \textbf{\#Corpus}  \\
         \midrule
    TREC-COVID          & Bio-Medical           & 50     & 171,332   \\
    NFCorpus           & Information           & 323    & 3,633     \\ \midrule
    NQ                 & Question               & 3,452 & 2,681,468   \\
    HotpotQA           & Answering              & 7,405  & 5,233,329 \\
    FiQA-2018          & (QA)                     & 648    & 57,638    \\ \midrule
    ArguAna            & Argument                  & 1,406  & 8,674     \\
    Touch\'e-2020      & Retrieval                 & 49     & 382,545   \\ \midrule
    CQADupStack        & Duplicate-Question     & 13,145 & 457,199   \\ \midrule 
    Quora              & Retrieval                 & 10,000 & 522,931   \\ \midrule
    DBPedia            & Entity-Retrieval      & 400    & 4,635,922 \\ \midrule
    SCIDOCS            & Citation-Prediction   & 1,000  & 25,657    \\ \midrule
    FEVER              & Fact Checking          & 6,666  & 5,416,568 \\ 
    Climate-FEVER      & Wikipedia             & 1,535  & 5,416,593 \\
    SciFact            & Scientific            & 300   & 5,183    \\
    \bottomrule
    \end{tabular}
    \label{tab:dataset_stats}
\end{table}

The retrieval effectiveness of \method{} is evaluated on the BEIR benchmark~\cite{thakur2021beir}, which includes 18 datasets that span a variety of domains.
Our evaluation focuses on the 14 publicly available datasets used for the zero-shot retrieval task. The statistics for these datasets are provided in Table~\ref{tab:dataset_stats}.

\textbf{Evaluation Metrics.}
To evaluate the retrieval effectiveness of \method{}, we use nDCG@10, the standard metric for the BEIR benchmark. The metric implementation follows the \texttt{pytrec-eval} toolkit~\cite{van2018pytrec_eval}, which is consistent with prior work~\cite{zhu2023large}.

\textbf{Baselines.}
We compare \method{} with several baseline retrievers implemented with different language models. GTR~\cite{ni2021large} employs large dual encoder-only models to build a dense retriever, while SGPT~\cite{muennighoff2022sgpt} trains dense retrieval models using decoder-only architectures. Emb-V3\footnote{\url{https://cohere.com/blog/introducing-embed-v3}} is a commercial text retrieval model provided by Cohere. PromptReps~\cite{zhuang2024promptreps} directly prompts LLMs to generate dense representations without supervision. RepLLaMA~\cite{ma2024fine} and E5-Mistral~\cite{wang2024multilingual} fine-tune LLMs as dense retrievers, using the hidden state of an additional end-of-sequence token to represent the input context. Notably, E5-Mistral is trained on the same dataset as \method{} but leverages a larger foundational model.
\begin{table*}[ht]
\caption{Overall Retrieval Performances on BEIR Benchmark. $^{\dagger}$ indicates the 11 most representative BEIR tasks used in CPT~\cite{neelakantan2022text} and Avg CPT Sub reflects the average performance across these tasks.} 
\centering
\resizebox{0.9\linewidth}{!}{
\begin{tabular}{l|c|c|c|c|c|c|c|cc}
    \toprule
    \textbf{Method} ($\rightarrow$) &\textbf{BM25}& \textbf{GTR} &  \textbf{SGPT} & \textbf{PromptReps} & \textbf{RepLLaMA} & \textbf{Emb-V3} & \textbf{E5-Mistral} & \multicolumn{2}{c}{\textbf{\method{}}} \\
    \textbf{Model Size} ($\rightarrow$) &/& 4.8B & 5.8B & 8B & 7B &/& 7B & 2.4B & 4B \\
    \midrule
    TREC-COVID$^{\dagger}$ &0.656& 0.501 & 0.873 & 0.693 & 0.847 & 0.794 & 0.708 & 0.795 & 0.836 \\
    NFCorpus$^{\dagger}$ &0.325& 0.342 & 0.362 & 0.330 & 0.378 & 0.336 & 0.353 & 0.378 & 0.399 \\
    NQ &0.329& 0.568 & 0.524 & 0.431 & 0.624 & 0.580 & 0.482 & 0.560 & 0.561 \\
    HotpotQA$^{\dagger}$ &0.603& 0.599 & 0.593 & 0.471 & 0.685 & 0.668 & 0.756 & 0.678 & 0.678 \\
    FiQA$^{\dagger}$ &0.236& 0.467 &0.372  & 0.324 & 0.458 & 0.388 & 0.545 & 0.434 & 0.462 \\
    ArguAna$^{\dagger}$ &0.414& 0.540 &0.514  & 0.330 & 0.486 & 0.508 & 0.625 & 0.567 & 0.562 \\
    Touché-2020$^{\dagger}$ &0.367 & 0.256 &0.254 & 0.218 & 0.305 & 0.319 & 0.191 & 0.211 & 0.250 \\
    Quora$^{\dagger}$ &0.789& 0.892 & 0.846 & 0.805 & 0.868 & 0.881 & 0.895 & 0.886 & 0.886 \\
    DBPedia$^{\dagger}$ &0.313& 0.408 &0.399  & 0.377 & 0.437 & 0.410 & 0.477 & 0.430 & 0.432 \\
    SCIDOCS &0.158& 0.161 &0.197 & 0.176 & 0.181 & 0.181 & 0.190 & 0.197 & 0.212 \\
    FEVER$^{\dagger}$ &0.753& 0.740 & 0.783 & 0.711 & 0.834 & 0.876 & 0.731 & 0.859 & 0.857 \\
    Climate-FEVER$^{\dagger}$ &0.213& 0.267 & 0.305& 0.214& 0.310 & 0.289 & 0.252 & 0.303 & 0.294 \\
    SciFact$^{\dagger}$ &0.665& 0.662 & 0.747 & 0.657 & 0.756 & 0.667 & 0.744 & 0.735 & 0.743 \\
    CQADupStack &0.299&0.399  & 0.381&/ & /&0.389&/&0.431&0.428\\
    \midrule
    \textbf{Avg CPT sub$^{\dagger}$} &0.485&0.516&0.550&0.466&0.579&0.558&0.571&0.571&0.582\\
    \textbf{Avg} &0.437& 0.486 & 0.511 &/  & / & 0.520 & / &0.533 & 0.543 \\
    \bottomrule
\end{tabular}}

\label{tab:overall}
\end{table*}

\textbf{Implementation Details.}
We initialize the \method{} models with MiniCPM-2.4B and MiniCPM-4B~\cite{hu2024minicpm}. All \method{} models are trained for 1,000 steps using the AdamW optimizer with a batch size of 256. The learning rate follows a cosine decay schedule, with a warm-up phase covering the first 3\% of the total iterations, peaking at 2e-4. We train \method{} using hybrid negatives, including one hard negative from the E5 dataset and seven in-batch negatives. The CoD length for all \method{} models is set to 8. \method{} is implemented using the OpenMatch toolkit~\cite{yu2023openmatch}, with flash-attention~\cite{NEURIPS2022_67d57c32} and LoRA~\cite{hu2021lora} enabled to mitigate memory constraints and improve computational efficiency. 

\section{Evaluation Results}

In this section, we first evaluate the retrieval effectiveness of
\method{} and then conduct ablation studies to show the roles of different modules in \method{}. Then we analyze the characteristics of learned embeddings during thinking step by step.

\subsection{Overall Performance}
The overall performance of \method{} and the baseline retrievers is shown in Table~\ref{tab:overall}.

Overall, \method{} outperforms all baseline retrievers in terms of average retrieval accuracy on BEIR, achieving more than a 2\% improvement. This highlights its effectiveness in enhancing the representation capability of LLMs for retrieval. Compared to the prompt-based method PromptReps, these fine-tuned LLM-based methods consistently show improvements, indicating that LLMs also benefit from supervised training to learn more tailored embeddings for retrieval.
When compared to E5-Mistral-7B, which is trained on the same E5 corpus as \method{}, \method{} significantly improves retrieval performance on TREC-COVID, NQ, and FEVER, demonstrating its capability across diverse question-answering scenarios.
Notably, when implemented with MiniCPM-2.4B, \method{} achieves retrieval performance comparable to that of larger 7B-scale LLM-based dense retrievers while utilizing only 35\% of the parameters. This demonstrates that \method{} can enhance the representation learning capabilities of smaller-scale LLMs, rather than relying on larger foundational LLMs. 
Furthermore, when implemented with MiniCPM-4B,  the retrieval performance of \method{} is improved by 1\%,  demonstrating that larger models effectively enhance the retrieval capabilities of \method{}.

\subsection{Ablation Study}

As shown in Table~\ref{tab:ablation}, we conduct ablation studies to further investigate the roles of Chain-of-Deliberation (CoD) and Self Distillation (SD) modules in \method{}.
\begin{table*}[h!]
    \centering
    \caption{Ablation Study of \textbf{D}\textbf{e}li\textbf{b}er\textbf{a}te \textbf{T}hinking based Dens\textbf{e} \textbf{R}etriever (\method{}). We train three \method{} variations: MiniCPM w/ SD, MiniCPM w/ CoD and vanilla MiniCPM. ${\dagger}$, ${\ddagger}$, and ${\mathsection}$ indicate statistically significant improvements over MiniCPM w/ SD$^{\dagger}$, MiniCPM w/ CoD$^{\ddagger}$ and vanilla MiniCPM$^{\mathsection}$.  }
    \resizebox{0.85\textwidth}{!}{
    \begin{tabular}{l|cccc|cccc}
        \toprule
        \multirow{2}{*}{\textbf{Method}} & \multicolumn{4}{c|}{\textbf{MiniCPM-2.4B}} & \multicolumn{4}{c}{\textbf{MiniCPM-4B}} \\
        & Vanilla & w/ SD & w/ CoD & \method{} & Vanilla & w/ SD & w/ CoD & \method{}\\
        \midrule
        TREC-COVID      & 0.728 &0.822 &0.805 &0.795 &0.747 &0.742 & 0.791&0.836 \\
        NFCorpus        & 0.368 &0.368 &0.371 &0.378 &0.379 &0.388 &0.378 &0.399 \\
        NQ              & 0.545 &0.531 &0.568 &0.560 &0.533 & 0.544      &0.508     &0.561 \\
        HotpotQA        & 0.670 &0.656 &0.669 &0.678 &0.564      &0.597      &0.631      &0.678 \\
        FiQA-2018       & 0.406 &0.409 &0.430 &0.434 &0.428 &0.428 &0.413 &0.462 \\
        ArguAna         & 0.561 &0.526 &0.547 &0.560 &0.569 &0.575 & 0.497&0.562 \\
        Touché-2020     & 0.202 &0.250 &0.219 &0.211 &0.195 &0.208 & 0.237&0.250 \\
        Quora           & 0.880 &0.788 &0.882 &0.886 &0.886 &0.890 &0.883 &0.886 \\
        SCIDOCS         & 0.191 &0.194 &0.195 &0.197 &0.210 &0.214 &0.198 &0.212 \\
        Climate-FEVER   & 0.277 &0.310 &0.258 &0.303 &0.211      &0.189      &0.184      &0.294 \\
        SciFact         & 0.715 &0.720 &0.733 &0.735 &0.731 &0.737 &0.730 &0.743 \\
        \midrule
        \textbf{Avg}    &0.504 &0.507&0.516 &0.522\rlap{$^{\dagger\mathsection}$} &0.496      & 0.501      &0.495     &0.535\rlap{$^{\dagger\ddagger\mathsection}$}\\
        \bottomrule
    \end{tabular}}

    \label{tab:ablation}
\end{table*}

We compare our \method{} with three variations, using MiniCPM-2.4B and MiniCPM-4B as the foundations for building dense retrievers. Both vanilla LLM and MiniCPM w/ CoD models represent documents using the hidden state of the last token and train query and document representations using contrastive training. The key difference between them lies in that MiniCPM w/ CoD performs additional CoD steps before obtaining the document representation. Besides, MiniCPM w/ SD is identical to \method{} but removes the CoD steps when generating the document representation.

Compared to vanilla LLM, MiniCPM w/ SD shows almost identical retrieval performance, indicating that relying solely on a few last tokens in the input sequence does not effectively enhance the document representations. This suggests that the special tokens used in CoD serve as prompts that stimulate LLMs to produce more meaningful embeddings.
On the other hand, MiniCPM w/ CoD still yields a limited improvement over the vanilla LLM, demonstrating that directly incorporating CoD in representing documents fails to enhance the representation ability of LLMs. After incorporating the Self Distillation mechanism, MiniCPM w/ CoD achieves further improvements, demonstrating its importance in capturing semantics from the different deliberative steps of CoD to optimize the last token as the document representation.
Additionally, when using contrastive training to optimize LLMs, the 4B-scale retrieval model performs worse than the 2.4B-scale model. Notably, \method{} not only mitigates this performance degradation but also leads to an additional 1.3\% improvement, highlighting the effectiveness and robustness of \method{}.

\subsection{Effectiveness of Chain-of-Deliberation with Different Thinking Depths}
\label{5_3_cod}
In this subsection, we explore how thinking depth affects the effectiveness of \method{}. Specifically, we vary the length of the Chain-of-Deliberation (CoD) to train several \method{}-2.4B and \method{}-4B models and evaluate their retrieval performance on TREC-COVID and FiQA. 

    

    

\begin{figure}[t]
    \centering
    \subfigure[TREC-COVID.]{ \label{fig:ndcg_treccovid}
    \includegraphics[width=0.48\linewidth]{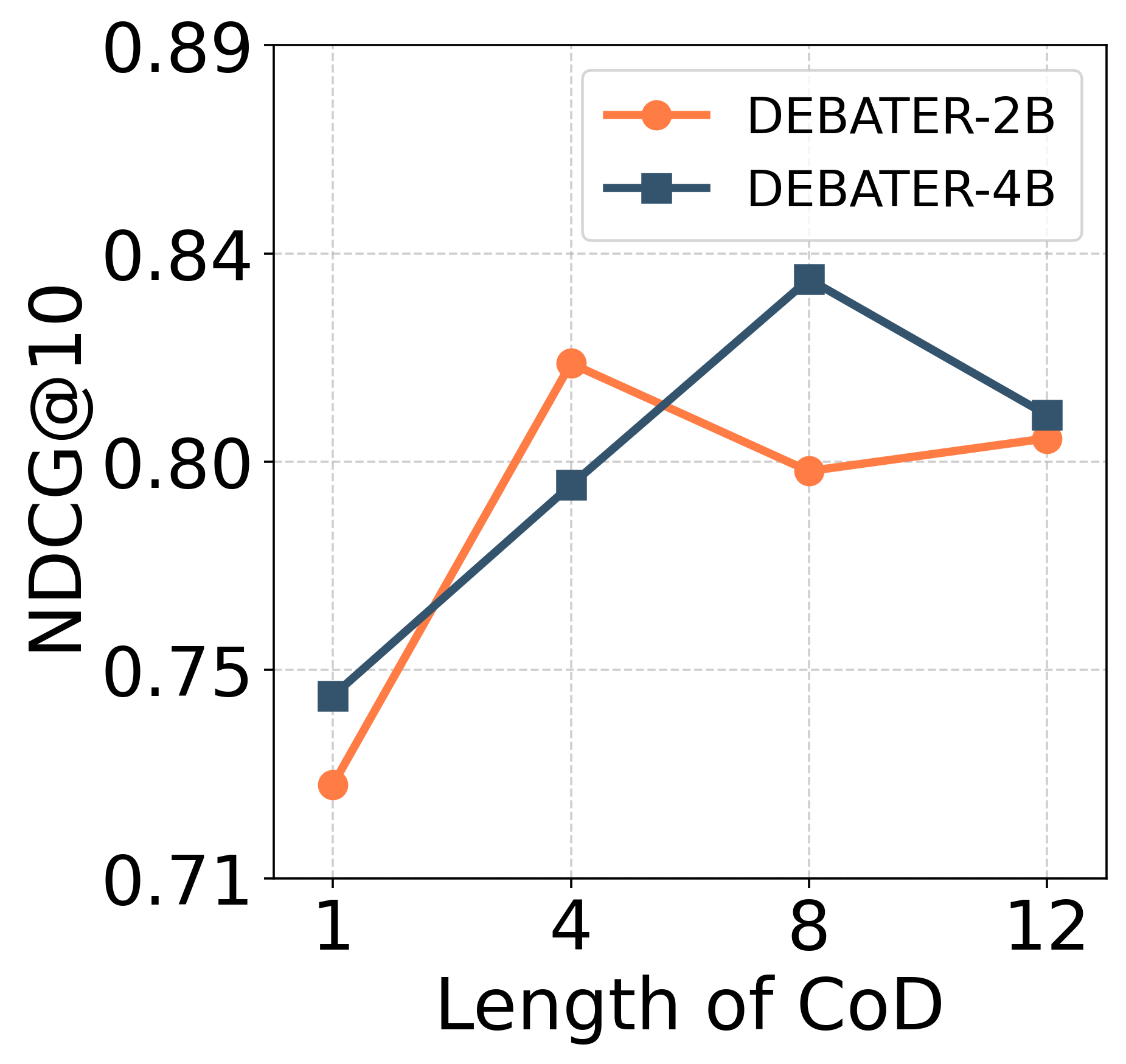}}
    \subfigure[FiQA.]{ \label{fig:fiqa}
    \includegraphics[width=0.48\linewidth]{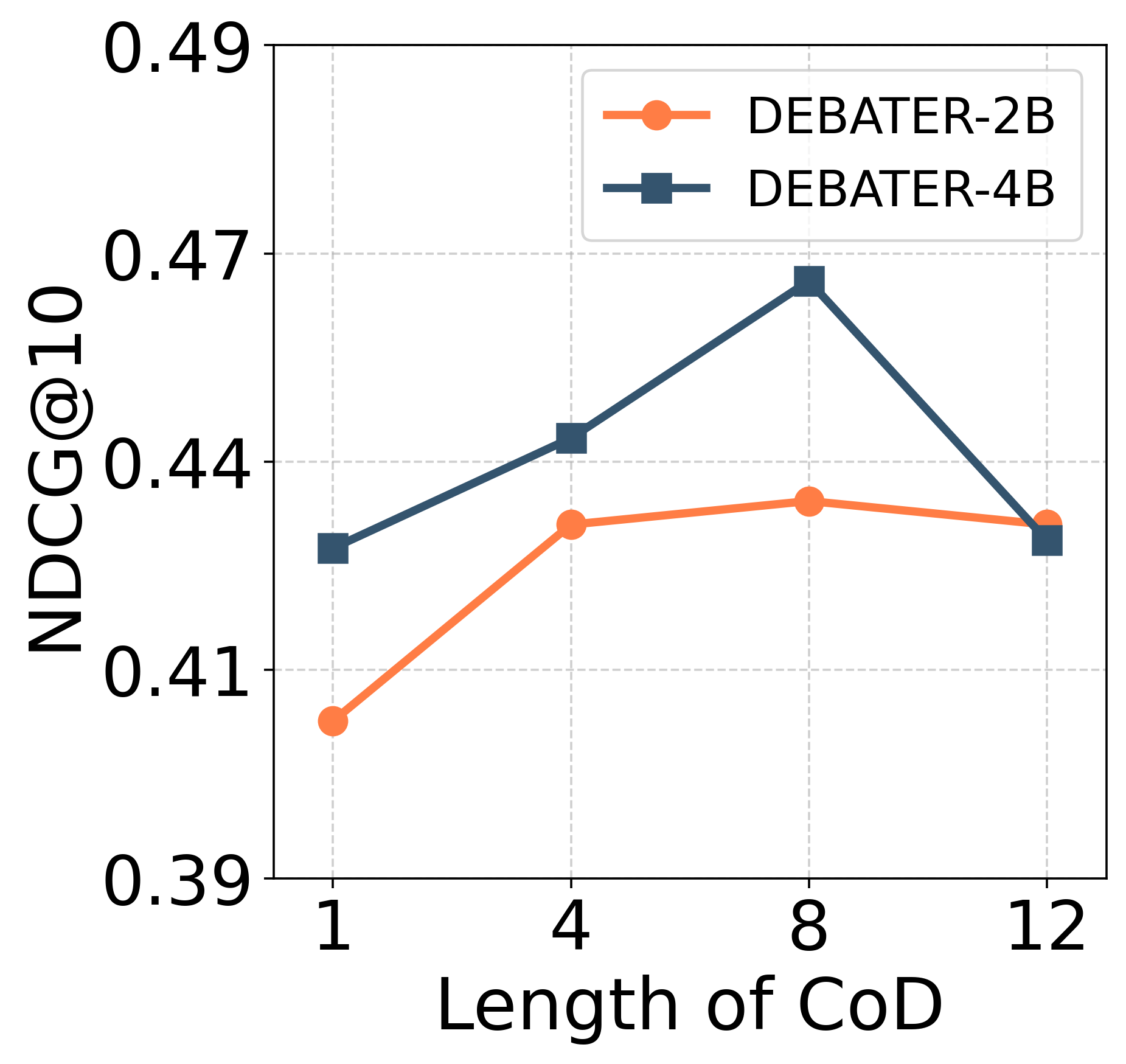}}
    \subfigure[NFCorpus.]{ \label{fig:ndcg_nfcorpus}
    \includegraphics[width=0.48\linewidth]{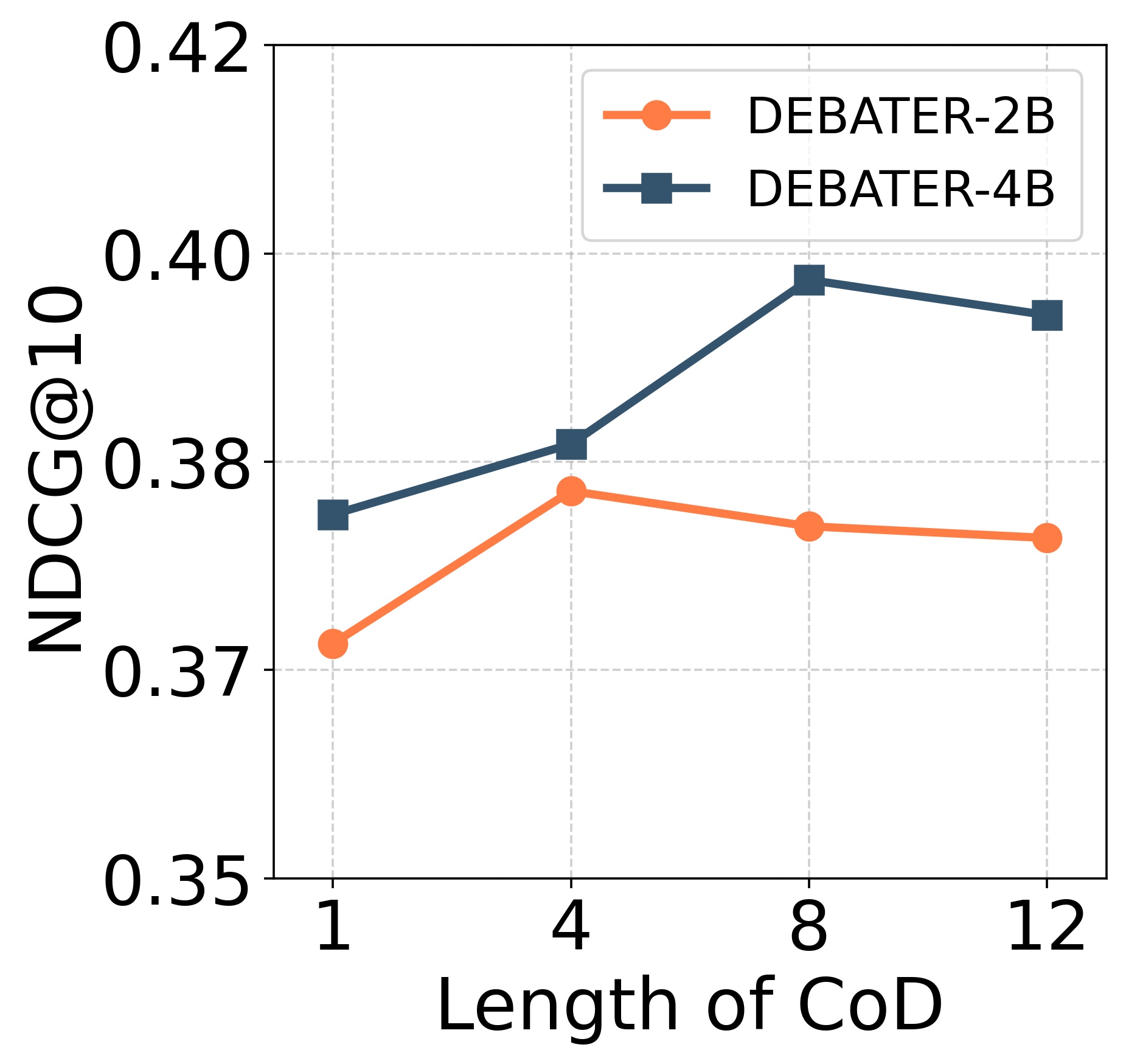}}
     \subfigure[SCIDOCS.]{ \label{fig:ndcg_scidocs}
    \includegraphics[width=0.48\linewidth]{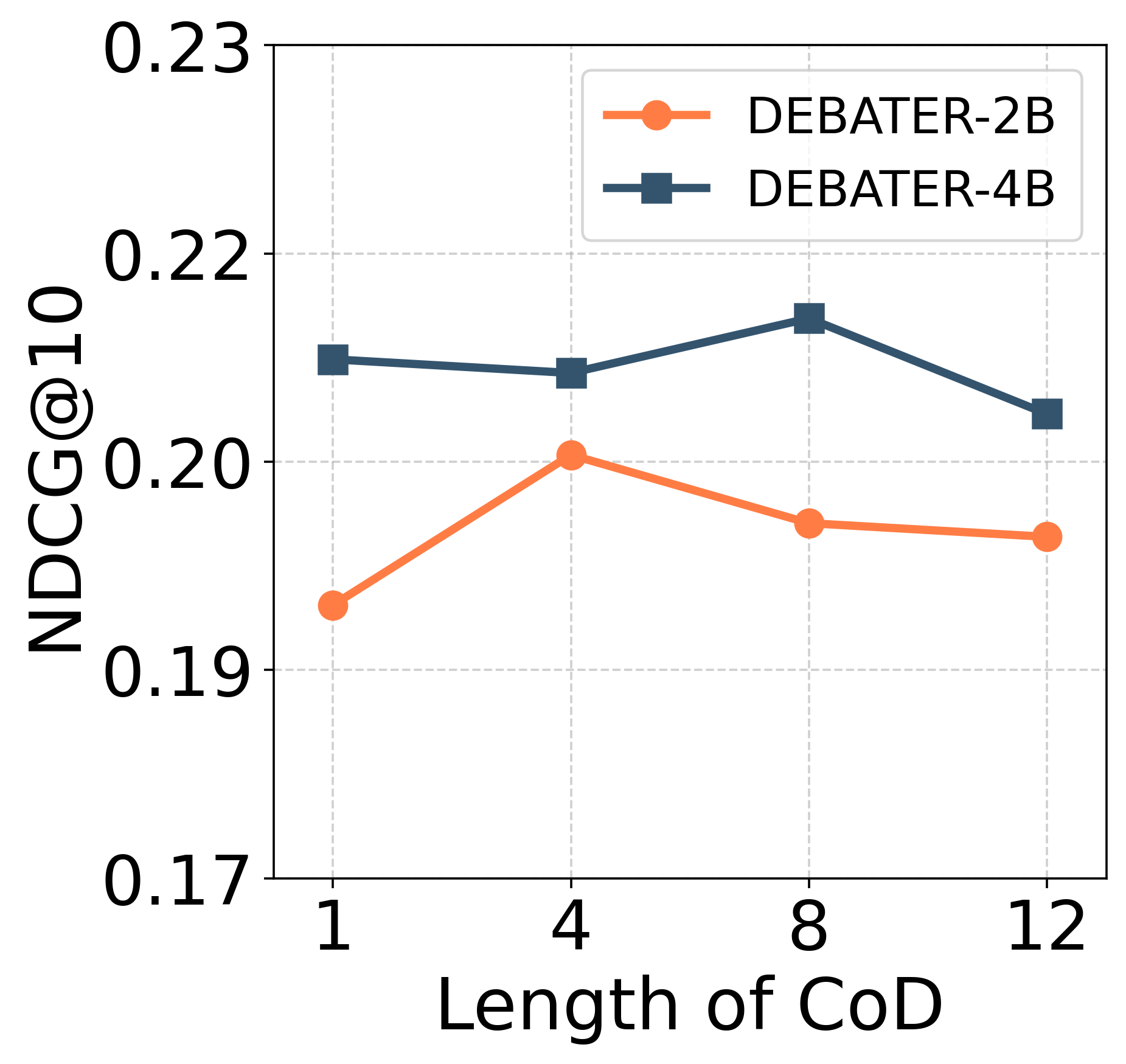}}
    \caption{Retrieval Performance of \method{} with Different Thinking Depths. We set the length of the CoD to train different \method{} models and evaluate them on different subsets of BEIR.}
    \label{fig:different_tokens}
\end{figure}
As illustrated in Figure~\ref{fig:different_tokens}, both \method{}-2.4B and \method{}-4B exhibit significant and consistent improvements in retrieval performance as the thinking depth increases to 4. 
This indicates that an appropriate thinking depth effectively activates the reasoning capabilities of LLM-based retrievers, enabling them to generate finer-grained representations of documents. 
When the thinking depth is further extended to 8, \method{}-2.4B reaches a plateau, indicating that it may be nearing its capacity to process more complex or prolonged deliberations.
In contrast, \method{}-4B continues to show incremental improvements when the length of CoD extends to 8, indicating that larger models benefit more from extended reasoning due to their stronger ability to integrate and retain detailed intermediate steps. 
Nonetheless, further increasing the CoD beyond a certain point (e.g., 12) may lead to overthinking and result in performance degradation for both model sizes. These observations demonstrate that while moderate depths effectively boost retrieval accuracy, excessively long chains can dilute the benefits and introduce unnecessary computational overhead. Overall, these findings underscore the importance of carefully tuning the thinking depth for LLM-based retrievers.

\begin{figure}[t]
    \centering
    \subfigure[TREC-COVID.]{ \label{fig:Treccovid-step}
    \includegraphics[width=0.485\linewidth]{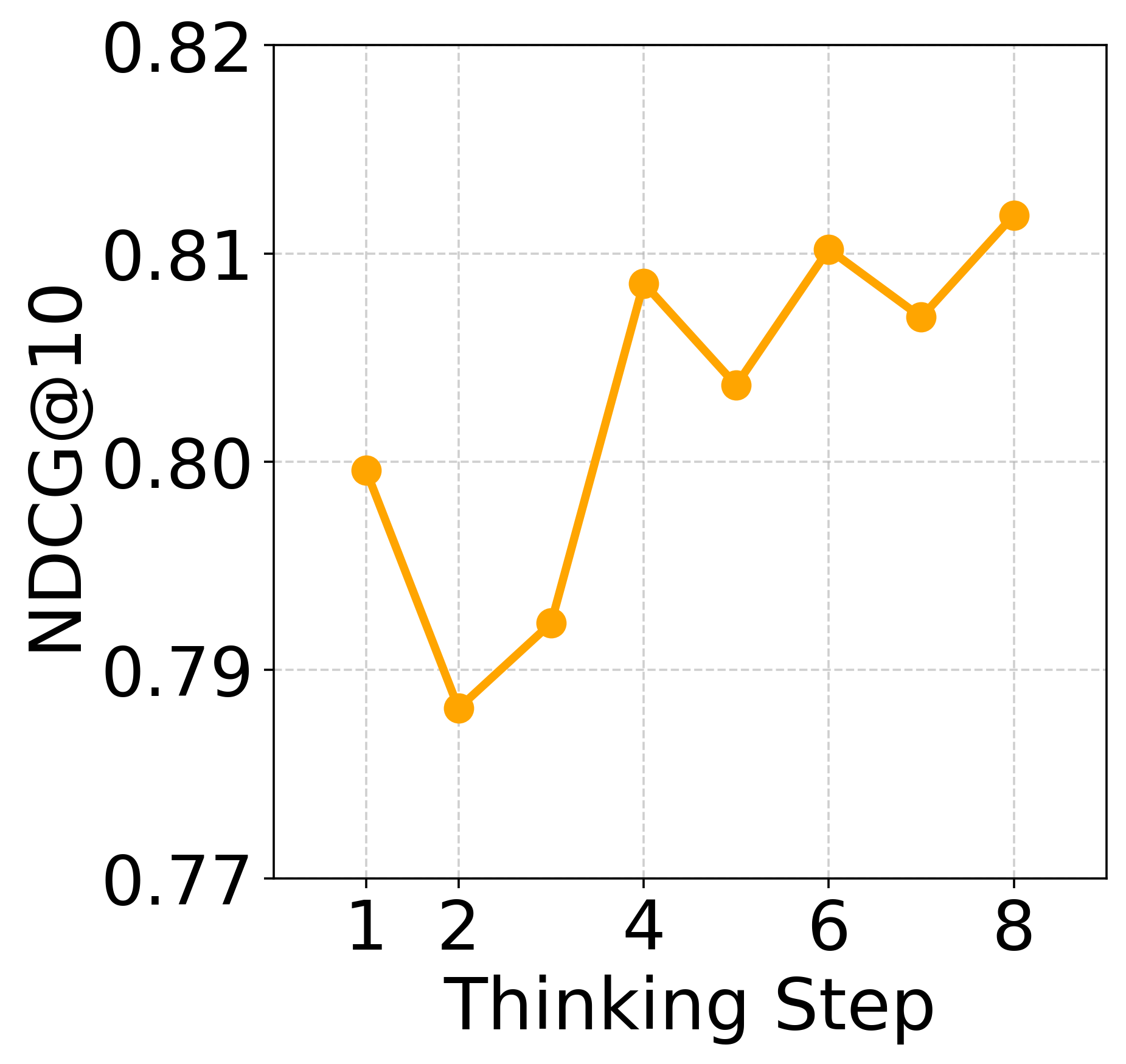}}
    \subfigure[FiQA.]{ 
    \label{fig:FIQA-step}
    \includegraphics[width=0.485\linewidth]{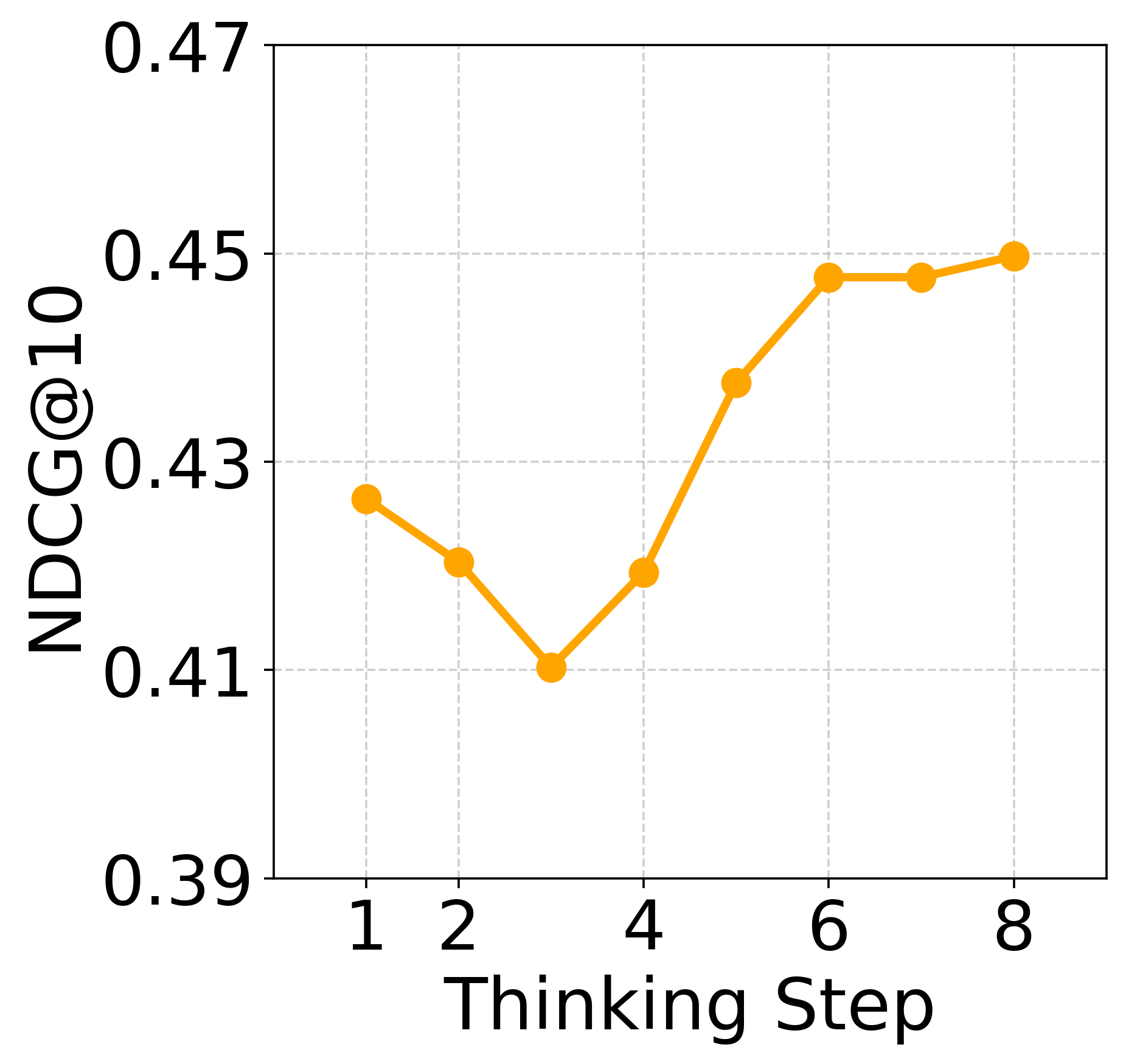}}
    \subfigure[NFCorpus.]{
    \label{fig:NF-step}
    \includegraphics[width=0.485\linewidth]{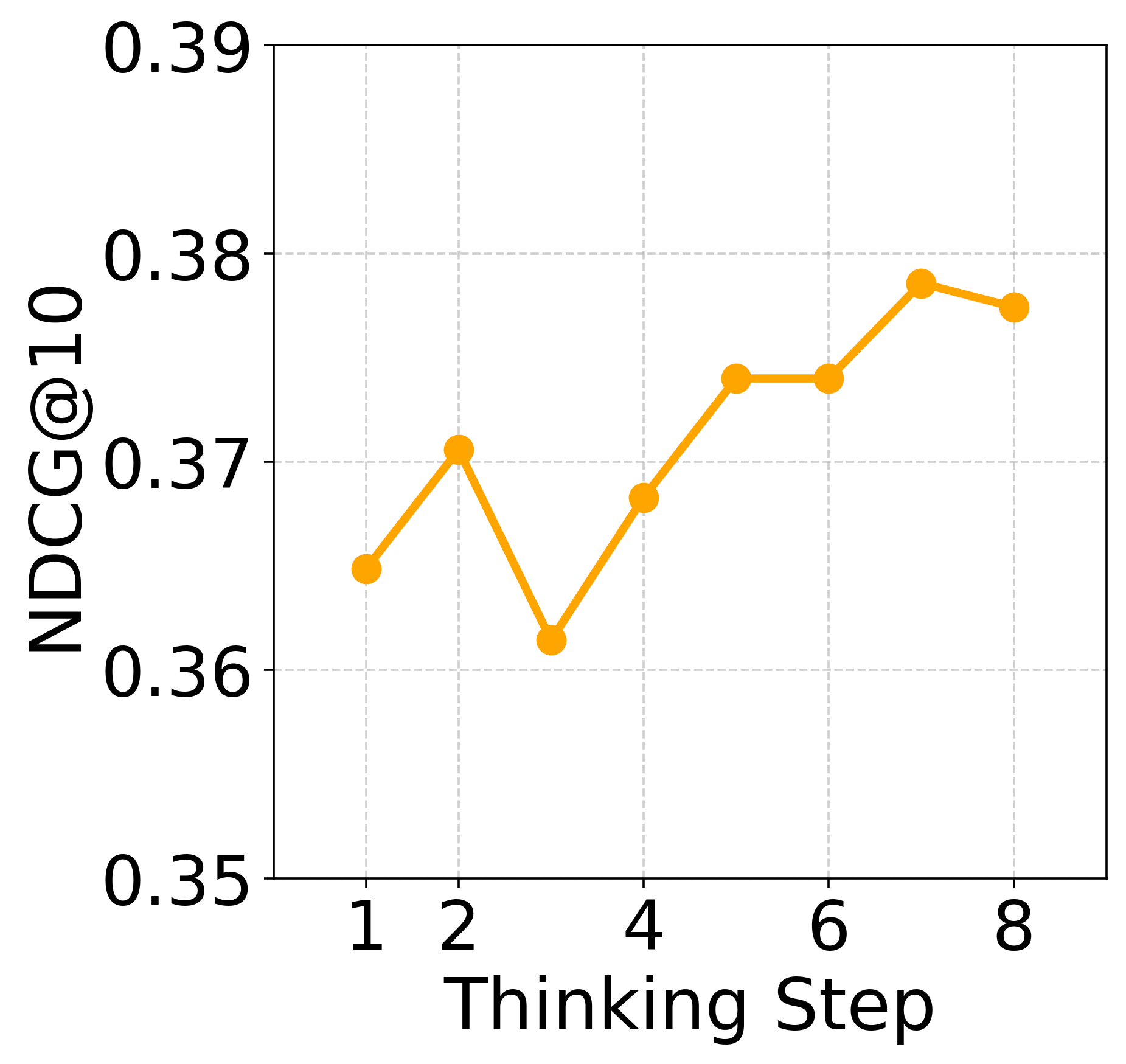}}
    \subfigure[SCIDOCS.]{
    \label{fig:SCId-step}
    \includegraphics[width=0.485\linewidth]{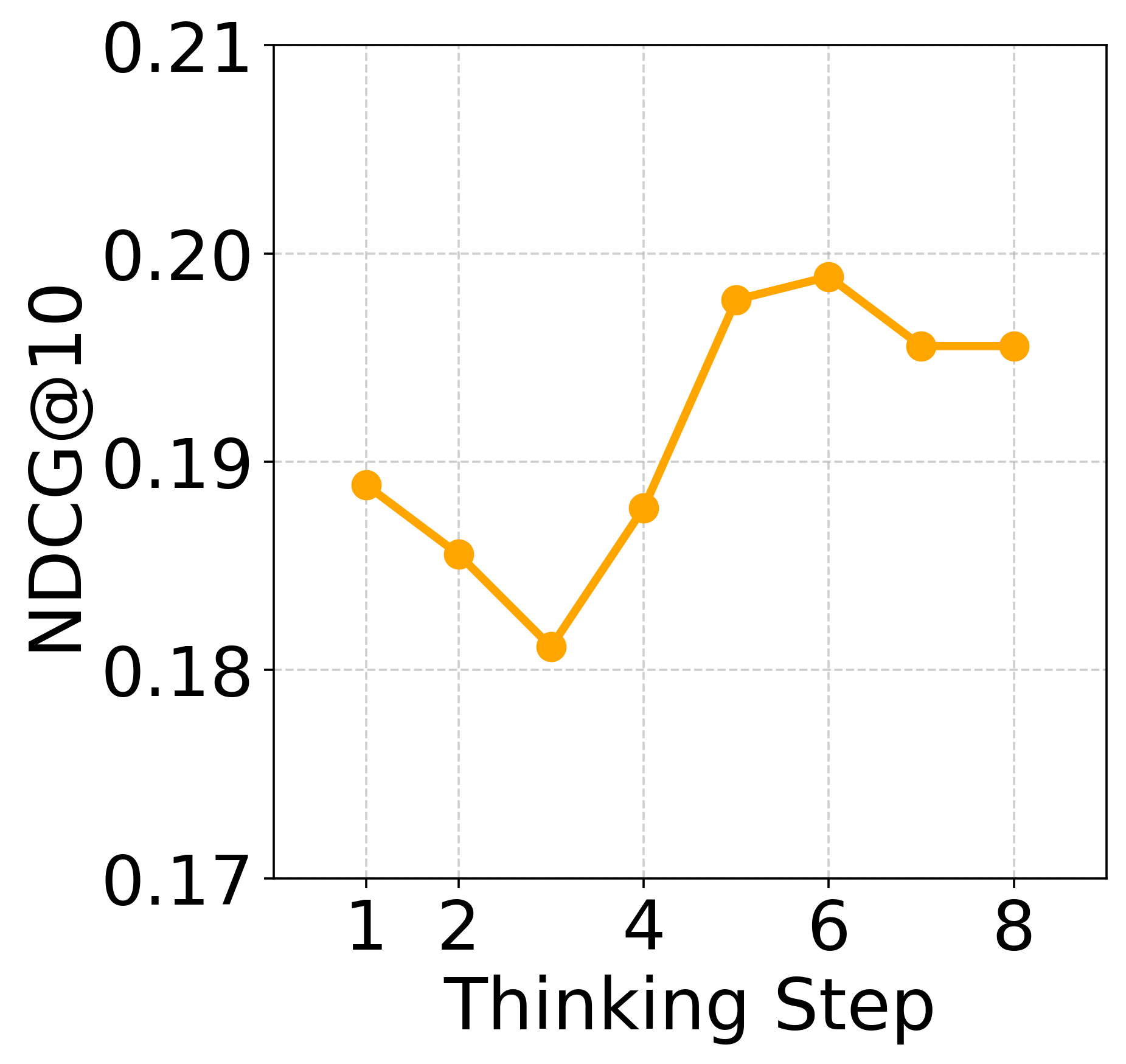}}
    \caption{Performance of \method{} at Different Thinking Steps. We collect all documents from each thinking step to demonstrate the retrieval performance of \method{} across different stages of reasoning.}
    \label{fig:think_step}
\end{figure}
\subsection{Retrieval Performance of CoD-Generated Document Representations} \label{5_4_CoD}

In this subsection, we investigate how the Chain-of-Deliberation (CoD) enhances the representation capability of LLM-based retrievers. Specifically, we evaluate the quality of embeddings produced at different thinking steps in CoD and assess their retrieval performance individually.

As shown in Figure~\ref{fig:think_step}, early steps (e.g., 1–2) produce relatively weak results, and performance may even drop.This suggests that initial embeddings, based on minimal deliberation, may lack the nuanced understanding required for effective retrieval. However, as the number of thinking steps increases, performance generally improves, indicating that more deliberation leads to more refined embeddings for retrieval. These results demonstrate that \method{} leverages the CoD mechanism to refine document representations step by step by reading information from previous steps. On the other hand, the gains eventually plateau, suggesting that once embeddings become sufficiently fine-grained, further deliberation provides limited benefits while increasing computational cost. 

\subsection{Characteristics of the Embeddings Generated by \method{}} 
In this subsection, we analyze the embeddings learned by \method{} from CoD. Specifically, we compute the average cosine similarity scores of embeddings generated at different positions in FiQA to understand how embeddings at various stages affect the final representation used for retrieval.

\begin{figure}[t]
    \centering
    \subfigure[\method{} w/o SD.]{ 
    \label{fig:hot_fig_wosdd}
    \includegraphics[width=0.48\linewidth]{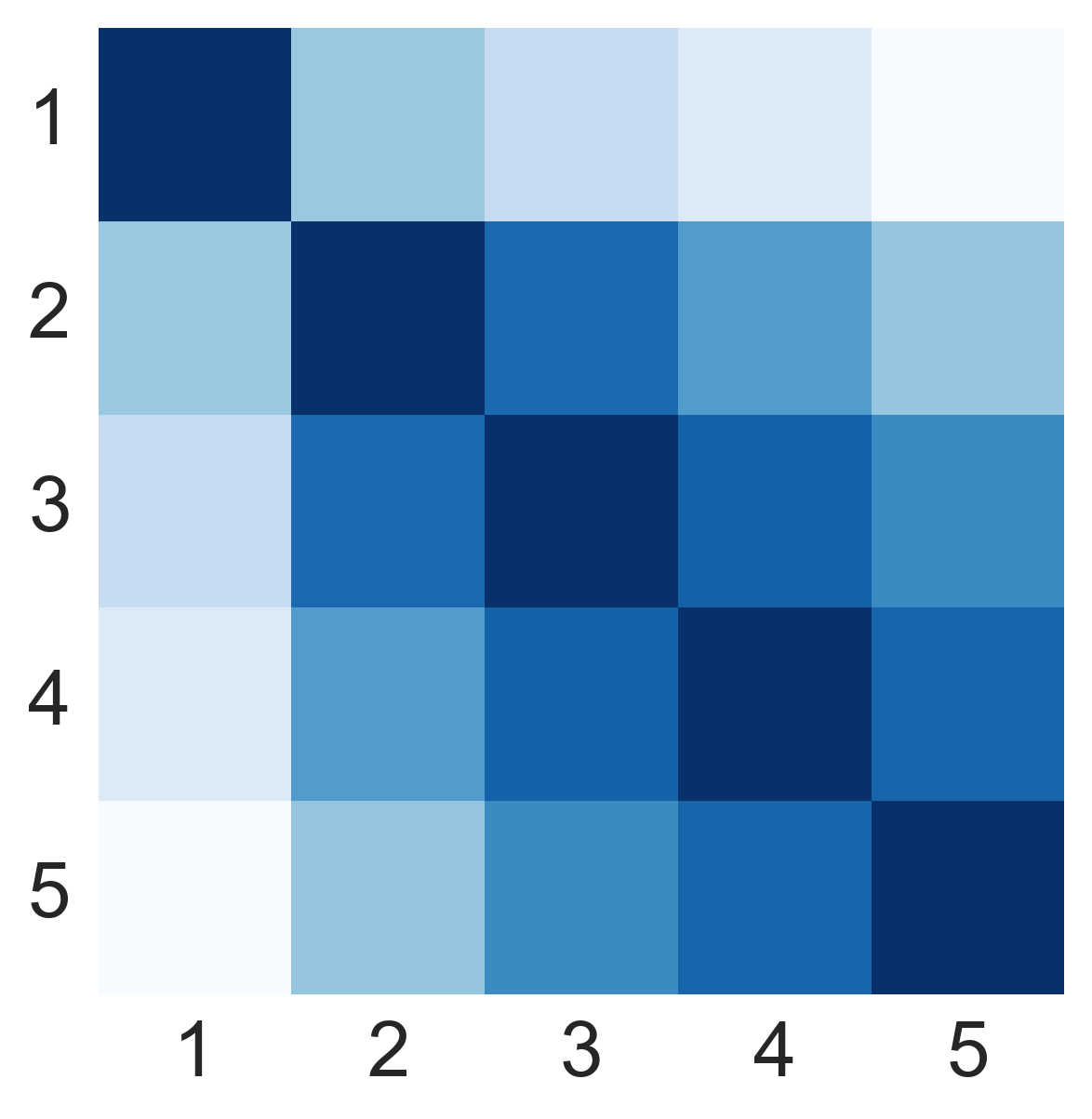}}
    \subfigure[\method{}.]{ 
    \label{fig:hot_fig}
    \includegraphics[width=0.48\linewidth]{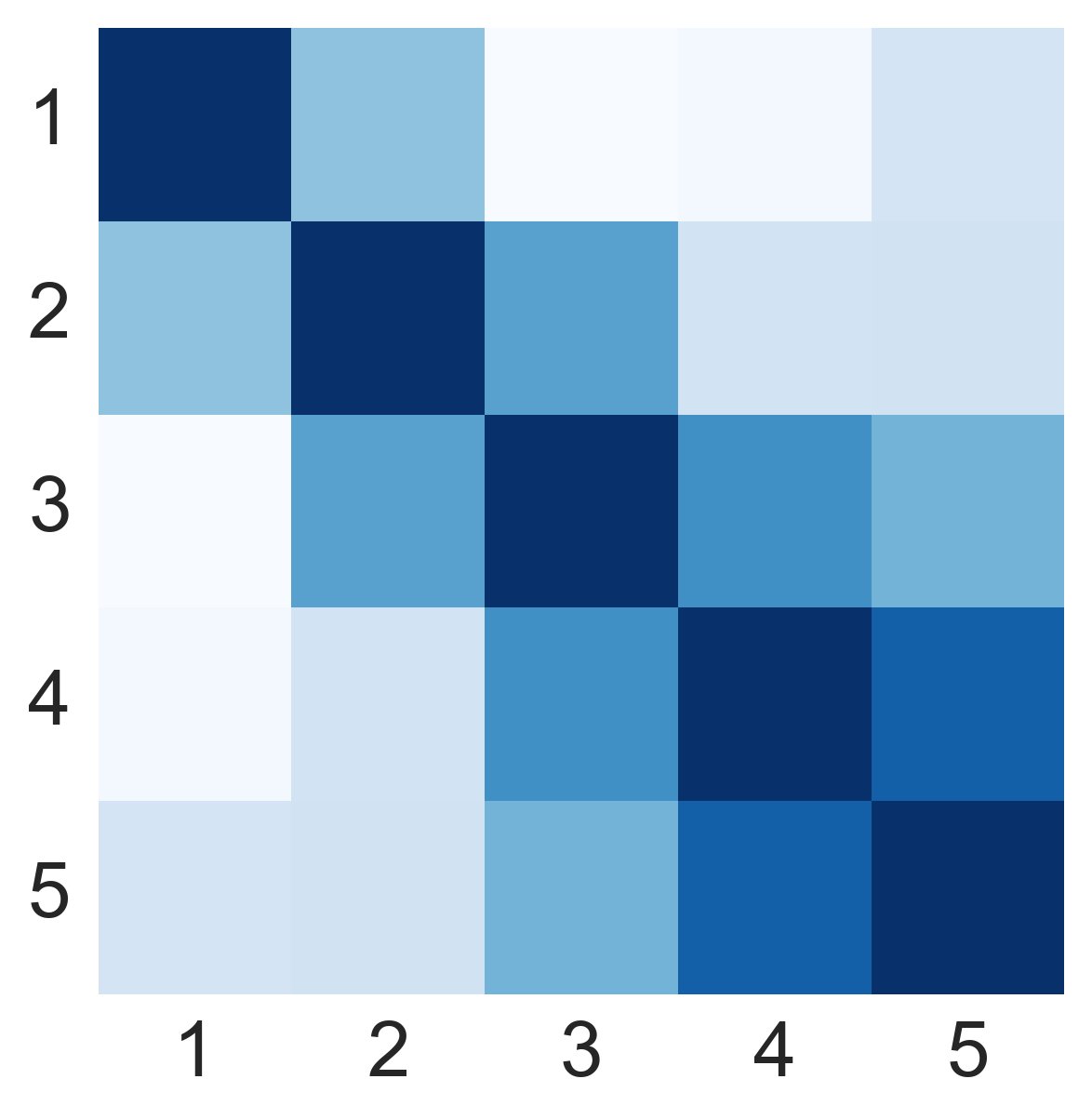}}
    \caption{Similarity Relationship Between Adjacent Position Embeddings. 
    Darker blue indicates a higher similarity score.}
    \label{fig:hot}
\end{figure}

\begin{figure}[t]
    \centering
    \subfigure[\method{} w/o SD.]{ \label{fig:sim_emb7_wosdd}
    \includegraphics[width=0.486\linewidth]{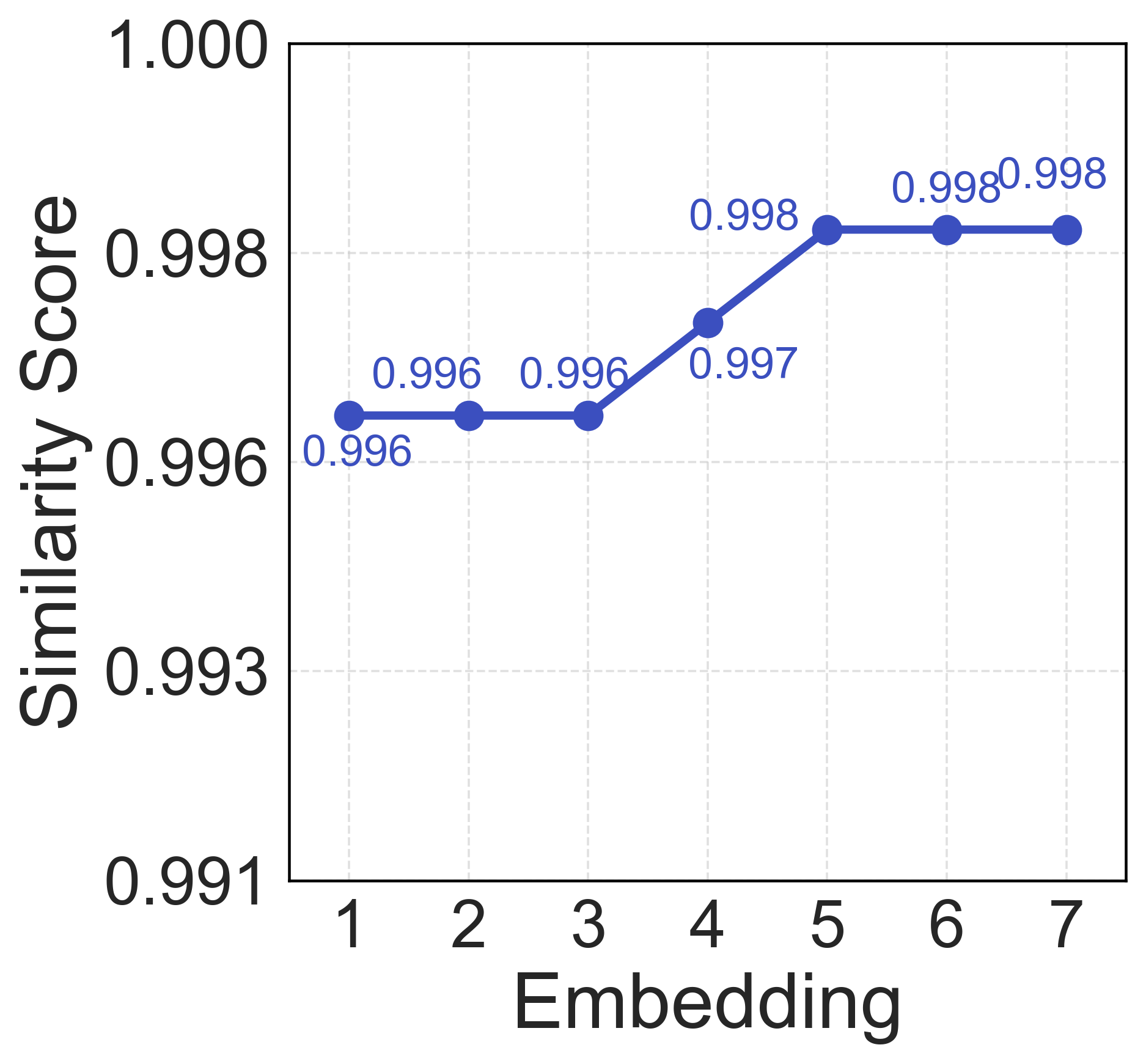}}
    \subfigure[\method{}.]{ \label{fig:sim_emb7}
    \includegraphics[width=0.486\linewidth]{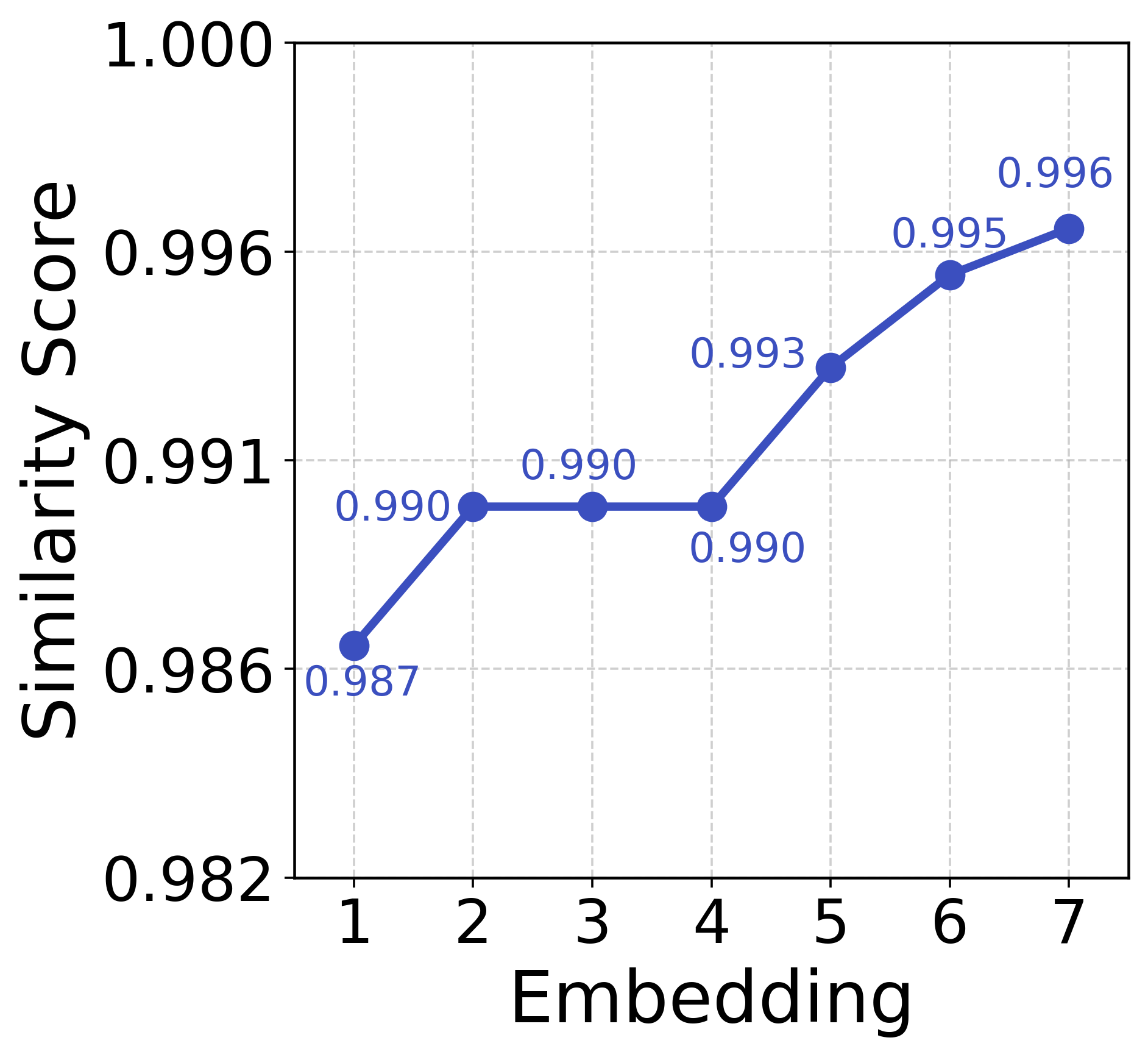}}
    \caption{Similarity Scores Between the First Seven Embeddings and the Last Embedding. The last embedding is used as the representation of documents for retrieval.}
    \label{fig:embed_similarity}
\end{figure}

\begin{table*}[t]
    \centering
    \caption{Retrieval Performance of \method{} across Different Model Scales. 
    }
    \resizebox{0.83\textwidth}{!}{
    \begin{tabular}{l|cc|cc|cc|cc}
        \toprule
        \multirow{2}{*}{\textbf{Method}} & \multicolumn{2}{c|}{\textbf{SmolLM2-135M}} & \multicolumn{2}{c|}{\textbf{SmolLM2-360M}}
        & \multicolumn{2}{c|}{\textbf{MiniCPM-1B}}
        & \multicolumn{2}{c}{\textbf{Qwen2.5-7B}}\\
        & Vanilla & \method{} & Vanilla & \method{}& Vanilla & \method{} & Vanilla & \method{}\\
        \midrule
        TREC-COVID      &0.681   &0.696 &0.721&0.646&0.678&0.778&0.770&0.829\\
        NFCorpus        &   0.286&0.299 &0.317&0.341&0.359&0.367&0.386&0.398\\
        FiQA-2018       & 0.223  &0.230 &0.292&0.378& 0.400&0.402&0.483&0.487\\
        ArguAna         &   0.455&0.428 &0.480&0.481&0.548&0.534&0.561&0.577\\
        Touché-2020     &  0.199&0.248 &0.187&0.208&0.162&0.227&0.242&0.235\\
        Quora           &0.841  &0.821 &0.869&0.874&0.884&0.883&0.891&0.893\\
        SCIDOCS         &0.135   &0.135 & 0.163&0.167&0.182&0.192&0.227&0.235\\
        SciFact         &0.530   &0.554&0.576&0.674&0.704&0.705&0.763& 0.767\\
        \bottomrule
        \textbf{Avg}   & 0.419&0.426&0.451&0.471&0.490&0.511&0.540&0.553\\
        \bottomrule
    \end{tabular}}
    \label{tab:small_model}
\end{table*}

\begin{table*}[t]
\centering
\caption{Case Studies. We present two cases from FiQA and TREC-COVID, and show the top 1 passage retrieved from MiniCPM and \method{}. Different colors are used to annotate important content: \sethlcolor{lightgreen}\hl{Blue} denotes critical information from the query, while \sethlcolor{lightred}\hl{Orange} highlights supporting details from the passage retrieved by \method{}.}

\resizebox{0.9\textwidth}{!}{
\begin{tabular}{lp{12.5cm}} 

\toprule
\multicolumn{1}{l}{\textit{\textbf{Case \#1 in  FiQA}}} \\
\midrule
\textbf{Query} & In the \sethlcolor{lightgreen}\hl{US}, is it a good idea to \sethlcolor{lightgreen}\hl{hire a tax consultant} for doing \sethlcolor{lightgreen}\hl{taxes}? \\   
\textbf{MiniCPM}: \textcolor{MyLightRed}{(Less Relevant)}&This may not exactly answer your question but, as a small business owner, I would highly recommend having a professional handle your taxes \textbf{...} I would recommend this especially if this is how you make your primary income, you can always write it off as a business expense.\\  
\textbf{\method{}}: \textcolor{red}{(Most Relevant)} & Whether you do decide to go with a tax advisor or not, be sure to do some research on your own. When we moved to the \sethlcolor{lightred}\hl{US} about 5 years ago, I did find the taxes here pretty complicated and confusing \textbf{···} After all, they are also humans prone to mistakes and your taxes are your liability in the end.  My suggestion is to \sethlcolor{lightred}\hl{start with a good tool that supports tax filing for non-residents}. Most of them provide a step-by-step QA based tool. As you go through the steps, Google each question you don't understand. \sethlcolor{lightred}\hl{It may take more time than hiring a tax advisor directly but in the end it will all be worth it.} \\
\midrule
\multicolumn{1}{l}{\textit{\textbf{Case \#2 in   TREC-COVID}}} \\
\midrule
\textbf{Query} &What is the \sethlcolor{lightgreen}\hl{mechanism of inflammatory response} and \sethlcolor{lightgreen}\hl{pathogenesis of COVID-19} cases?\\ 
\textbf{MiniCPM}: \textcolor{MyLightRed}{(Less Relevant)} &The novel coronavirus disease (COVID-19) pandemic is placing significant strains on health systems 
\textbf{...} In this context, the worlds scientific biomedical establishment is unleashing an unprecedented response to the COVID-19 pandemic.
In this commentary, based on a very recent research report, we intend to highlight how a new mechanism describing the RAGE transactivation produced by Ang II-mediated ATR1 activation can run continuously and thus, reinforcing a sustained inflammation in lungs, due to the SARS-Cov-2-mediated imbalance of the ACE/And II/ATR1 pathway.\\  

\textbf{\method{}}: \textcolor{red}{(Most Relevant)} &The evidence on the pathophysiology of the novel coronavirus SARS-CoV-2 infection is rapidly growing \textbf{...} 
The answer to this question would allow rationalizing the fear surrounding this pandemic. Understanding of \sethlcolor{lightred}\hl{the pathophysiology of COVID-19 relies on an understanding of interplaying mechanisms}, including \sethlcolor{lightred}\hl{SARS-CoV-2 virulence}, \sethlcolor{lightred}\hl{human immune response}, and \sethlcolor{lightred}\hl{complex inflammatory reactions with coagulation} playing a major role 
\textbf{...} More importantly, a comprehensive understanding of pathological mechanisms of COVID-19 will increase the efficacy of therapy and decrease mortality. Herewith, presented is the current state of knowledge on COVID-19:  \sethlcolor{lightred}\hl{beginning from the virus}, \sethlcolor{lightred}\hl{its transmission}, and \sethlcolor{lightred}\hl{mechanisms of entry into the human body}, through the pathological effects on the cellular level, up to \sethlcolor{lightred}\hl{immunological reaction}, \sethlcolor{lightred}\hl{systemic and organ presentation.} Last but not least, currently available and possible future therapeutic and diagnostic options are briefly commented on.\\
\bottomrule
\end{tabular}}
\label{tab:case_study}
\end{table*}

\textbf{Learning Patterns of CoD.} As shown in Figure~\ref{fig:hot_fig}, we present the average similarity scores among the first five embeddings generated by \method{} to explore how \method{} refines document representations step by step during CoD.

The results reveal a clear pattern in the similarity relationships: each embedding is most similar to its immediate neighbors, with similarity gradually decreasing as the distance between embeddings increases. This indicates that each embedding heavily relies on the previously decoded representations to generate more refined embeddings, which likely results from the autoregressive decoding mechanism of LLMs. Comparing this with the \method{} w/o SD model (Figure~\ref{fig:hot_fig_wosdd}), we observe that the \method{} model shows higher similarity scores with representations from more recent steps during CoD. This suggests that our Self Distillation method effectively encourages LLMs to learn more diverse representations at different thinking steps and to gather more relevant information from nearby steps, which leads to finer-grained document representations. 

\textbf{Contributions of CoD Steps to Document Representations.} 

Figure~\ref{fig:embed_similarity} illustrates the similarity relationship between embeddings at intermediate thinking steps of CoD and the final document representation generated at the last thinking step. This helps us explore the contributions of different thinking steps to the final document representations.

In general, both \method{} w/o SD and \method{} models exhibit a trend of gradually increasing similarity to the final embedding as the thinking steps progress. As shown in Figure~\ref{fig:sim_emb7_wosdd}, the \method{} w/o SD model tends to produce similar similarity scores with the final step, showing that relying solely on CoD may degrade the performance of \method{}. It may lie in that all CoD generated embeddings are supervised with the same training loss and optimized to match the same query, making them become homogeneous. In contrast, \method{} (Figure~\ref{fig:sim_emb7}) shows a more significant increase in similarity, indicating that these thinking steps contribute more variably to the final document representation. Notably, the information generated at each CoD step is gradually compressed into the last embedding, which further demonstrates the effectiveness of \method{} in leveraging the thinking capacity of LLMs to generate more effective document representations for retrieval. 


\subsection{Effectiveness of \method{} across Different Model Scales}

In this subsection, we examine whether the proposed \method{} remains effective across language models of varying scales. 

We evaluate \method{} on a range of language models with varying parameter scales, including SmolLM2-135M, SmolLM2-360M~\cite{allal2025smollm2smolgoesbig}, MiniCPM-1B, and Qwen2.5-7B-Instruct~\cite{qwen2025qwen25technicalreport}. These models span from lightweight architectures suitable for edge deployment to more advanced large language models. Due to limited computational resources, our experiments were conducted on a subset of low-resource datasets from the BEIR benchmark, which still provide a diverse set of retrieval tasks for evaluation.  

As shown in Table~\ref{tab:small_model}, the results demonstrate that \method{} consistently improves average performance across diverse retrieval tasks for all model sizes. Notably, even with the smallest model, SmolLM2-135M, \method{} delivers performance gains on most tasks, indicating its effectiveness under limited capacity.  Moreover, as model size increases, the performance improvements become more pronounced, reflecting \method{}’s ability to better leverage the increased thinking capacity and representational power of larger models. These findings highlight the robustness and scalability of \method{} across LLMs of different scales.

\subsection{Case Studies}

In this subsection, we present two case studies on FiQA and TREC-COVID to illustrate the effectiveness of \method{}. Table~\ref{tab:case_study} shows the top-1 retrieved documents from MiniCPM and \method{}.

For the FiQA query “In the US, is it a good idea to hire a tax consultant for doing taxes?”, MiniCPM retrieves a generic document suggesting hiring a professional but ignores the explicit “US” context. In contrast, \method{} retrieves a document that fully aligns with the query, including the US-specific aspect. This shows that \method{} enables finer-grained representations for contextually appropriate retrieval.

For the TREC-COVID query “What is the mechanism of inflammatory response and pathogenesis of COVID-19 cases?”, MiniCPM returns documents with many pathological terms but lacking clear explanations of the mechanisms, reflecting limited fine-grained understanding. \method{}, however, retrieves results that directly address the mechanisms, demonstrating the benefit of deliberate reasoning before retrieval.

\section{Conclusion}
This paper proposed the \textbf{D}\textbf{e}li\textbf{b}er\textbf{a}te \textbf{T}hinking based Dens\textbf{e} \textbf{R}etriever (\method{}), a novel method designed to enhance the reasoning capabilities of LLM-based dense retrievers via deliberation-augmented embedding. 
Through the integration of Chain-of-Deliberation (CoD) and Self Distillation (SD), \method{} significantly improves retrieval performance by capturing different views of documents before generating final embeddings. 
Our experimental results demonstrate that \method{} outperforms existing dense retrievers by implementing with the LLM of a smaller scale.

\begin{acks}
This work is partly supported by the National Natural Science Foundation of China (No. 62206042) and the Fundamental Research Funds for the Central Universities (No. N25ZLL045). This work is also supported by the AI9Stars community.
\end{acks}

\bibliographystyle{ACM-Reference-Format}
\balance
\bibliography{custom}

\end{document}